\documentclass[11pt,a4paper]{article}
\pdfoutput=1
\usepackage{jheppub}
\usepackage{amsfonts}
\usepackage{amsmath}
\usepackage{cancel}
\usepackage{graphicx}
\usepackage{hyperref}
\usepackage[latin1]{inputenc}
\usepackage{multirow}
\usepackage[table]{xcolor}
\usepackage{xspace}

\newcommand{\HEPfit}{\texttt{HEPfit}\xspace}

\hypersetup{pdfstartview=FitV,colorlinks=true,linkcolor=blue,citecolor=red,filecolor=black,urlcolor=blue}

\definecolor{color1}{HTML}{00dd00}
\definecolor{color2}{HTML}{dd0000}

\definecolor{color8TeV}{HTML}{ddddff}
\definecolor{color13TeV}{HTML}{eeee99}

\begin{document}

\title{Constraints on coloured scalars from global fits}

\author[a]{Otto Eberhardt,}
\emailAdd{otto.eberhardt.physics@gmail.com}   %%% otto.eberhardt@ific.uv.es

\author[a]{V\'ictor Miralles,}
\emailAdd{victor.miralles@ific.uv.es}

\author[a]{Antonio Pich}
\emailAdd{antonio.pich@ific.uv.es}

\affiliation[a]{Instituto de F\'isica Corpuscular, Parque Cient\'ifico, 
C/Catedr\'atico Jos\'e Beltr\'an, 2, E-46980 Paterna, Spain}

\abstract{
We consider a simple extension of the electroweak theory, incorporating one $SU(2)_L$ doublet of
colour-octet scalars with Yukawa couplings satisfying the principle of minimal flavour violation.
Using the HEPfit package, we perform a global fit to the available data, including all relevant theoretical constraints, and extract the current bounds on the model parameters. Coloured scalars with masses below 1.05 TeV are already excluded, provided they are not fermiophobic. The mass splittings among the different (charged and CP-even and CP-odd neutral) scalars 
are restricted to be smaller than 20 GeV. Moreover, for scalar masses smaller than 1.5 TeV, the Yukawa coupling of the coloured scalar multiplet to the top quark cannot exceed the one of the SM Higgs doublet by more than 80\%. These conclusions are quite generic and apply in more general frameworks (without fine tunings). The theoretical requirements of perturbative unitarity and vacuum stability enforce relevant constraints on the quartic scalar potential parameters that are not yet experimentally tested.
}

\preprint{{\raggedleft IFIC/21-24 \par}}

\setcounter{tocdepth}{1}

\maketitle

%%%%%%%%%%%%%%%%%%%%%%%%%%%%%%%%%%%%%%%%%%%%%%%%%%%%%
\section{Introduction}
\label{sec:intro}
%%%%%%%%%%%%%%%%%%%%%%%%%%%%%%%%%%%%%%%%%%%%%%%%%%%%%

The main focus of the first LHC run was on the search for the Standard Model (SM) Higgs boson and culminated in its spectacular discovery in 2012 \cite{Aad:2012tfa, Chatrchyan:2012xdj}. After the conclusion of the run 2, the properties of this new scalar are much better known, and they seem to agree very well with the SM predictions. In combination with direct LHC searches for additional spin-$0$ particles, they put strong bounds on extensions of the scalar sector of the SM, as for example the Two-Higgs-Doublet model \cite{Eberhardt:2020dat}. On the other hand, new physics beyond the Standard Model (BSM) is needed to fix the shortcomings of the current theoretical framework, and many BSM scenarios could manifest themselves through the appearance of additional scalars at the LHC.
Most of these SM extensions contain scalar fields which transform as singlets under the $SU(3)_C$ group of QCD and, therefore, do not carry colour charge. But coloured scalars are less constrained than one might naively think: they do not mix with the SM Higgs and contribute to its signal strengths only at the loop level. However, recent LHC searches for scalar resonances are expected to yield strong constraints on their properties.

Here we want to study a simple extension of the SM scalar sector with one $SU(2)_L$ doublet of scalar  fields that transforms as an octet under the colour $SU(3)_C$ group. In some sense, one can think of it as a Two-Higgs-Doublet model in which one of the doublets is coloured. Scalars of this type
with masses around the electroweak scale could emerge naturally from $SU(5)$ or $SO(10)$ unification theories at high energy scales \cite{Georgi:1979df,Dorsner:2006dj,Perez:2016qbo,Bertolini:2013vta}.
As this model was first proposed by Manohar and Wise \cite{Manohar:2006ga}, we will refer to it as the Manohar-Wise (MW) model in the following. Several phenomenological aspects of this dynamical scenario have been studied in the literature \cite{Gresham:2007ri, Gerbush:2007fe, Burgess:2009wm,Degrassi:2010ne, He:2011ti, Dobrescu:2011aa, Bai:2011aa, Arnold:2011ra, Kribs:2012kz, Reece:2012gi, Cao:2013wqa, He:2013tla, Cheng:2015lsa,Cheng:2016tlc, Martinez:2016fyd, Hayreter:2017wra,Cheng:2018mkc,Hayreter:2018ybt,Gisbert:2019ftm,Miralles:2019uzg,Miralles:2020tei}, 
yet a comprehensive analysis including all relevant up-to-date constraints has not been performed.

In this article we will present such an analysis. After defining the MW model in Section \ref{sec:model}, we briefly describe in Section~\ref{sec:constraints} the \HEPfit package \cite{deBlas:2019okz} that we employ to perform the global fits. This section explains in detail the various constraints on the parameter space of the model, which come from theoretical considerations (Section \ref{sec:theoryconstraints}), Higgs measurements  (Section \ref{sec:Higgsconstraints}), direct LHC searches (Section \ref{sec:searchconstraints}) and flavour observables (Section \ref{sec:Flavourconstraints}). 
Putting all pieces together, we show the results of our fits in Section \ref{sec:results}. Finally, Section \ref{sec:summary} contains a summary of our findings and a brief outlook on future prospects.

%%%%%%%%%%%%%%%%%%%%%%%%%%%%%%%%%%%%%%%%%%%%%%%%%%%%%
\section{The scalar colour-octet model}
\label{sec:model}
%%%%%%%%%%%%%%%%%%%%%%%%%%%%%%%%%%%%%%%%%%%%%%%%%%%%%
The MW model extends the SM with one electroweak doublet of colour-octet scalar fields with hypercharge $Y=\frac{1}{2}$. Since it has colour charge, the new scalar multiplet does not mix with the SM Higgs doublet. Furthermore, the coloured particles cannot acquire a vacuum expectation value (vev) because colour must be conserved. Therefore, only the SM Higgs boson will acquire a vev which will minimise the most general potential that can be built with this scalar sector:
\begin{align}
 V_{\text{\tiny{MW}}}
 & = m_\Phi^2\Phi^\dagger\Phi + \tfrac12 \lambda \left(\Phi^\dagger\Phi \right)^2 + 2 m_S^2 {\rm Tr}\left(S^{\dagger i} S^{\phantom{\dagger}}_i\right)+ \mu_1 {\rm Tr}\left(S^{\dagger i} S^{\phantom{\dagger}}_i S^{\dagger j} S^{\phantom{\dagger}}_j\right) 
  + \mu_2 {\rm Tr}\left(S^{\dagger i} S^{\phantom{\dagger}}_j S^{\dagger j} S^{\phantom{\dagger}}_i\right)  \nonumber  
  \\
 &\phantom{{}={}}
 + \mu_3 {\rm Tr}\left(S^{\dagger i} S^{\phantom{\dagger}}_i\right) {\rm Tr}\left(S^{\dagger j} S^{\phantom{\dagger}}_j\right)
  + \mu_4 {\rm Tr}\left(S^{\dagger i} S^{\phantom{\dagger}}_j\right) {\rm Tr}\left(S^{\dagger j} S^{\phantom{\dagger}}_i\right) 
  + \mu_5 {\rm Tr}\left(S^{\phantom{\dagger}}_i S^{\phantom{\dagger}}_j\right) {\rm Tr} \left(S^{\dagger i} S^{\dagger j}\right) 
   \nonumber \\
 &\phantom{{}={}}
  + \mu_6 {\rm Tr}\left(S^{\phantom{\dagger}}_i S^{\phantom{\dagger}}_j S^{\dagger j} S^{\dagger i}\right)+ \nu_1 \Phi^{\dagger i}\Phi_{i} {\rm Tr}\left(S^{\dagger j} S^{\phantom{\dagger}}_j\right) 
  + \nu_2 \Phi^{\dagger i}\Phi_{j} {\rm Tr}\left(S^{\dagger j} S^{\phantom{\dagger}}_i\right) \nonumber \\
 &\phantom{{}={}}
 +\left[ \nu_3 \Phi^{\dagger i}\Phi^{\dagger j} {\rm Tr}\left(S^{\phantom{\dagger}}_i S^{\phantom{\dagger}}_j\right) 
           +\nu_4 \Phi^{\dagger i} {\rm Tr}\left(S^{\dagger j} S^{\phantom{\dagger}}_j S^{\phantom{\dagger}}_i\right) 
           +\nu_5 \Phi^{\dagger i} {\rm Tr}\left(S^{\dagger j} S^{\phantom{\dagger}}_i S^{\phantom{\dagger}}_j\right) 
           + {\rm h.c.}\right] , \label{eq:genpot}
\end{align}
where $\Phi = (\phi^+,\phi^0)^T$ is the usual SM doublet, 
the traces are taken in colour space, and  $i$ and $j$ denote $SU(2)_L$ indices. The additional $(8,2)_{1/2}$ scalar fields
$S^A =(S^{A,+},S^{A,0})^T$ are contained in the multiplet $S=S^AT^A$ with $T^A$ the generators of the $SU(3)_C$ group.
All potential parameters are real except $\nu_3$, $\nu_4$ and $\nu_5$, but performing a phase rotation we can always take $\nu_3$ to be also real. 

From Eq.~\eqref{eq:genpot} we can derive the masses of the physical neutral octet scalar ($m_R$), the neutral octet pseudoscalar ($m_I$) and the charged octet ($m_{S^\pm}$), which are split by the Higgs vev, $\langle \phi^0\rangle = v/\sqrt{2}$:

\begin{equation}
    m^2_{S^\pm}=m^2_S+\nu_1 \,\frac{v^2}{4}\, ,
    \qquad \qquad
    m^2_{R,I}=m^2_S+(\nu_1+\nu_2\pm 2\, \nu_3)\, \frac{v^2}{4}\, .
    \label{eq:msplit}
\end{equation}

The kinetic term (the factor two gives the correct canonical normalisation for the fields)
\begin{equation}\label{eq:kin}
\mathcal{L}_{K}\, =\, 2\,\mathrm{Tr}[(D_\mu S)^\dagger D^\mu S]
\end{equation}
generates the interaction of the octet scalars with the gauge bosons through the covariant derivative
\begin{equation}
D_\mu S\, =\, \partial_\mu S + i g_s\, [G_\mu,S] + i g\,\frac{\sigma^i}{2}\, W_\mu^i S
+\frac{i}{2}\, g' B_\mu S \, ,
\end{equation}
with $G_\mu = G_\mu^A T^A$ the octet gluon field.
Thus, these interactions are determined by the gauge symmetry and do not introduce additional free parameters.

The coloured scalars can also couple to the quarks through the Yukawa interaction. In order to guarantee the suppression of unwanted flavour-changing neutral currents, which are extremely suppressed experimentally, we will assume the principle of Minimal Flavour Violation (MFV) \cite{Chivukula:1987py,DAmbrosio:2002vsn}, which is based on the hypothesis that all Yukawa matrices are proportional to the same flavour structures that break the $SU(3)_{Q_L}\otimes SU(3)_{u_R}\otimes SU(3)_{d_R}$ symmetry in the SM. This is in fact one of the main motivations of the MW model because the only scalar representations that can couple to quarks and be compatible with this principle are the colour octet or singlet electroweak doublets \cite{Manohar:2006ga}.
With this assumption, the Yukawa couplings of the coloured scalars take the form
\begin{align}
 {\cal L}_Y \supset -\sum^3_{i,j=1} \left[\eta_D Y^d_{ij}\,\bar{Q}_{L_i}S d_{R_j}+\eta_U Y^u_{ij}\,\bar Q_{L_i}\tilde{S}u_{R_j} + \text{h.c.}\right] .
 \label{eq:LY}
\end{align}
Here, $i$ and $j$ are family indices, $Y^q = \sqrt{2} M_q/v$ ($q=u,d$) denote the up and down
SM Yukawa matrices and the tilde in the $S$ field indicates charge conjugation. The proportionality constants $\eta_U$ and $\eta_D$ are, in general, complex parameters.

Looking at Eqs.~\eqref{eq:genpot} and \eqref{eq:LY}, we observe that the MW model contains 18 more parameters than the SM, 14 of which are real while the other 4 are imaginary phases. In order to simplify the phenomenological analysis and to reduce the total number of free parameters we will only work in the CP-conserving limit. This assumption removes the imaginary parts of $\nu_4$, $\nu_5$, $\eta_U$ and $\eta_D$ and we end up with only 14 new free parameters.

%%%%%%%%%%%%%%%%%%%%%%%%%%%%%%%%%%%%%%%%%%%%%%%%%%%%%
\section{Fit constraints}
\label{sec:constraints}
%%%%%%%%%%%%%%%%%%%%%%%%%%%%%%%%%%%%%%%%%%%%%%%%%%%%%

Our statistical data analysis will be based on a global Bayesian fit. 
We make use of the public HEPfit package \cite{deBlas:2019okz}, which is interfaced with the Bayesian Analysis Toolkit \cite{Caldwell:2008fw}. This code has been already applied to several BSM analyses, including the Two-Higgs-Doublet model \cite{Cacchio:2016qyh,Gori:2017tvg,Chowdhury:2017aav,Eberhardt:2020dat} and the Georgi-Machacek model \cite{Chiang:2018cgb}. We have adapted the code, including the additional routines needed to study the MW model.
These routines are also public and can be extended in future works to incorporate additional observables. In our fit we have only included observables that have been directly calculated with this model. Constraints obtained from existing bounds on higher-dimension operators\footnote{See for instance Ref. \cite{Farina:2018lqo} (and references therein) where some particular colour-octet models are analysed. Owing to the MFV assumption which strongly suppresses the scalar couplings to light quarks, the $pp\to t\bar t$ analysis of Ref. \cite{Farina:2018lqo} is not relevant for the MW model.}
are omitted, since we prefer  to directly include in the fit the observables used to derive those constraints. Indeed, HEPfit has also been proven to be extremely reliable to constrain higher-dimension operators \cite{deBlas:2018tjm, Durieux:2019rbz, Miralles:2021dyw}.
One of the key features of HEPfit is its independence of other codes at runtime, which provides a very fast framework for statistical analyses.

 Since we will try to use in our fit all the available information, we did not have any previous constraints on the MW parameters. Therefore we decided to use a uniform distribution as a prior for the 14 free parameters of the MW model. The ranges adopted are shown in Table~\ref{tab:parameters}. The range taken for $\eta_U$ follows from the assumption of a perturbative top Yukawa coupling, while larger values of $\eta_D$ are possible since $m_b\ll m_t$. The dependence of our results on the priors used turns out to be small, as long as these priors are reasonable. For instance, increasing the range of $m_S^2$ up to $2^2$ TeV$^2$ leads to the same limits, but we do not gain any further information since many of the direct-searches analyses included do not go beyond $1.5$ TeV. For $\nu_n$, $\mu_n$ and $\eta_U$ the posterior probabilities are basically zero in the limits of the ranges considered and the same constraints are always found if this (reasonable) condition is satisfied. The only exception is $\eta_D$ which could not be constrained within the range considered. However, higher values of $\eta_D$ will always bring stronger limits on the other parameters. 
The parameter space is constrained by the theoretical and experimental requirements that we now discuss.

\begin{table}[h!]
\begin{center}
\begin{tabular}{|c|c|c|c|c|c|}
\hline
Parameters&$\textcolor{black}{m_S^2}$&$\textcolor{black}{\nu_n}$&$\textcolor{black}{\mu_n}$&$\textcolor{black}{\eta_U}$&$\textcolor{black}{\eta_D}$\\
\hline
Priors &$\textcolor{black}{(0.4^2, 1.5^2)}$ TeV$\textcolor{black}{^2}$&(-10, 10)&(-10, 10)&(-5, 5)&(-20, 20)\\
\hline
\end{tabular}
\caption{Prior values of the fitted parameters.\label{tab:parameters}}
\end{center}
\end{table}

%%%%%%%%%%%%%%%%%%%%%%%%%%%%%%%%%%%%%%%%%%%%%%%%%%%%%
\subsection{Theoretical constraints}
\label{sec:theoryconstraints}
%%%%%%%%%%%%%%%%%%%%%%%%%%%%%%%%%%%%%%%%%%%%%%%%%%%%%

Since we are interested in the LHC phenomenology, we will assume that the MW model is well defined up to the few TeV scale. As a consequence, the renormalisation group (RG) evolution should be stable between the electroweak scale and 5 TeV. By RG stability we mean the absence of Landau poles as well as the fact that the scalar potential $V_{\text{\tiny{MW}}}$ should be bounded from below at any of the considered scales. The RG equations (without Yukawa and gauge couplings) were taken from Ref.~\cite{He:2013tla} and a set of necessary positivity bounds was derived in Ref.~\cite{Cheng:2018mkc}.

Another strong constraint on the quartic couplings of $V_{\text{\tiny{MW}}}$ is the unitarity requirement that the two-to-two scattering processes of the scalar particles should not have a probability larger than $1$. This condition is usually expressed in terms of the partial wave amplitudes $a_j$. Using their LO ($a_j^{(0)}$)  and NLO ($a_j^{(1)}$) expressions \cite{Cheng:2018mkc}, perturbative unitarity can be imposed at LO, NLO and at the so-called NLO+ approximation \cite{Murphy:2017ojk} that includes the square of the NLO correction and therefore contains some, but not all, NNLO terms \cite{Cheng:2018mkc}: 
\begin{equation}\label{eq:NLLO+}
    \left| a_j^{(0)} + \mathrm{Re}\!\left(a_j^{(1)}\right)\right|^2\,\le\,\frac{1}{4}\, .
\end{equation}
Note also that the LO and NLO expressions for the partial wave amplitudes are only available in the large--$s$ approximation and should not be applied below a certain scale $\mu_u$, which we choose to be $\mu_u=1.5$ TeV as the current limits on the mass of the colour-octet scalars are around 1 TeV \cite{Miralles:2019uzg}.

On top of these unitarity bounds, we also impose perturbative behaviour of the quantum corrections and allow only for scenarios in which the NLO contributions to the partial wave amplitudes are smaller in magnitude than the LO term.

In the fits, we run up to 3 or 5 TeV each parameter set, sampled at the electroweak scale,  and control at each intermediate step if the parameters still comply with positivity and (above $\mu_u$) perturbative unitarity. If this is not the case, the running is stopped and the corresponding cut-off scale is returned as an output.

%%%%%%%%%%%%%%%%%%%%%%%%%%%%%%%%%%%%%%%%%%%%%%%%%%%%%
\subsection{Higgs constraints}
\label{sec:Higgsconstraints}
%%%%%%%%%%%%%%%%%%%%%%%%%%%%%%%%%%%%%%%%%%%%%%%%%%%%%

\begin{table}[h!]
{\small
\begin{center}
\begin{tabular}{|c |c| c c c| c c c|}
\hline
\multicolumn{1}{|c}{\textbf{Production}} & \textbf{Decay} & \textbf{Reference} & \textbf{$L$} & \textbf{$\sqrt{s}$} & \textbf{Reference} & \textbf{$L$} & \textbf{$\sqrt{s}$}\\[3pt]
\multicolumn{1}{|c}{}& & & \textbf{[fb$^{-1}$]} & \textbf{[TeV]} & & \textbf{[fb$^{-1}$]} & \textbf{[TeV]} \\[3pt]
\cline{3-8}

\multicolumn{1}{|c}{}& &\multicolumn{3}{c|}{\textbf{ATLAS}}&\multicolumn{3}{c|}{\textbf{CMS}}\\
\hline
ggF & \multirow{ 4}{*}{$h \to \gamma\gamma$} & \multirow{ 2}{*}{\cite{ Aad:2014eha}}   & \multirow{ 2}{*}{4.5/20.3}  & \multirow{ 2}{*}{7/8} & \multirow{ 2}{*}{\cite{ Khachatryan:2014ira}} & \multirow{ 2}{*}{5.1/19.7} & \multirow{ 2}{*}{7/8}\\
VBF &&&&&&& \\
VH &  & \multirow{ 2}{*}{\cite{ATLAS:2020pvn}}& \multirow{ 2}{*}{139} & \multirow{ 2}{*}{13}  & \multirow{ 2}{*}{ \cite{Sirunyan:2021ybb}} & \multirow{ 2}{*}{137} & \multirow{ 2}{*}{13} \\
ttH &  & &  &  &  & & \\
\hline
\multirow{ 10}{*}{ggF} & \multirow{ 2}{*}{$\mu^+ \mu^-$}  & - & - & - & \cite{ Khachatryan:2016vau} & 5/20& 7/8 \\
   &  & \cite{Aad:2020xfq}  &  139  & 13 & \cite{Sirunyan:2020two}   & 137 & 13 \\\cline{2-8}
  &  \multirow{ 2}{*}{$\tau^+ \tau^-$ } & \cite{Aad:2015vsa}& 4.5/20.3 & 7/8 & \cite{Chatrchyan:2014nva}  & 4.9/19.7 & 7/8 \\
     &   &\cite{ATLAS:2020qdt} & 139 & 13&  \cite{CMS:2020gsy}  & 137 &13 \\\cline{2-8}
  &   \multirow{ 2}{*}{$WW$} & \cite{ ATLAS:2014aga, Aad:2015ona}  & 25, 4.5/20.3 & 7/8 & \cite{Chatrchyan:2013iaa} & 4.9/19.4& 7/8 \\
     &  & \cite{ATLAS:2020qdt} &  139  & 13 &  \cite{CMS:2020gsy} &  137 & 13 \\\cline{2-8}
  &  \multirow{ 2}{*}{$Z\gamma$ } & \cite{Aad:2015gba} & 4.7/20.3 & 7/8 &\cite{Chatrchyan:2013vaa}  & 5/19.6 & 7/8 \\
     &  & \cite{Aad:2020plj} & 139 & 13 & \cite{Sirunyan:2018tbk}  & 35.9 & 13 \\\cline{2-8}
  &   \multirow{ 2}{*}{$ZZ$} & \cite{Aad:2014eva}&  4.5/20.3& 7/8 & \cite{Khachatryan:2014jba} & 5.1/19.7 & 7/8 \\
   &  & \cite{Aad:2020mkp} & 139 & 13 &\cite{CMS:2020gsy} & 137 & 13 \\\hline

\end{tabular}
\caption{Higgs signal strengths measured by ATLAS and CMS.}
\label{Tab:HiggsStrengths}
\end{center}
}
\end{table}

%%%%%%%%%%%%%%%%%%%%%%%%%%%%%%%%%%%%
%%%%%%%%%%%%%%%%%%%%%%%%%%%%%%%%%%%%
%%%%%%%%%%%%%%%%%%%%%%%%%%%%%%%%%%%%

In the MW model there are additional contributions to some of the production and decay channels of the SM Higgs boson, due to the presence of the colour-octet scalars. At the LHC, the SM Higgs boson is produced mainly through gluon fusion (ggF), vector boson fusion (VBF), associated production with a $t\bar{t}$ pair (tth) and associated production with vector bosons (Wh/Zh). The LO contribution to the first process is a one-loop amplitude already in the SM. As the coloured scalars couple to gluons and to the SM Higgs boson, they will contribute to the gluon-fusion production mode  at the same order than the SM. Similarly, the decay of the SM Higgs boson to photons is also a one-loop process in the SM to which the new scalars contribute.

The LHC data for the Higgs physics are parametrized  in terms of the Higgs signal strengths, which are defined as the measured cross section times branching ratio for a given production and decay Higgs channel, in units of the SM prediction. Table~\ref{Tab:HiggsStrengths} compiles the experimental papers from which we have taken the values of the different Higgs signal strengths. Comparing the measured values of these observables with their theoretical predictions 
\cite{Cheng:2016tlc}, we will try to constrain the parameters $\nu_1$, $\nu_2$ and $\nu_3$, which are directly related with the mass splittings of the new scalars in Eq. \eqref{eq:msplit}.

%%%%%%%%%%%%%%%%%%%%%%%%%%%%%%%%%%%%%%%%%%%%%%%%%%%%%
\subsection{Constraints from direct searches}
\label{sec:searchconstraints}
%%%%%%%%%%%%%%%%%%%%%%%%%%%%%%%%%%%%%%%%%%%%%%%%%%%%%

In order to find constraints from direct experimental searches of additional scalars, we will compare the theoretical prediction of the production cross-section times branching ratio of a given process, $\sigma\cdot\mathcal{B}$, with the experimental upper limits of the ATLAS and CMS collaborations. The channels that we have included in the fit are shown in Table~\ref{tab:directsearches1}. Since decays of the colour-octet scalars into purely electroweak bosons are forbidden, we have only considered their decays into gluons and quarks. Furthermore, in order to be sure that the decay modes studied are indeed the dominant ones, we impose for each search that the mass difference between the coloured scalar we are analysing and the lightest member of the $S$ multiplet is smaller than the $W^\pm$ mass \cite{Miralles:2019uzg}. In this way, we forbid kinematically the decays into another coloured scalar and a gauge boson, which can become the dominant decay modes for some configurations of the parameter space. As we will show later, this assumption is well justified, given the constraints that perturbative unitarity enforce on the mass splittings of the new scalars.

\begin{table}[h!]
\begin{center}
\begin{small}
\begin{tabular}{|l|lc|c|c|}
\hline
\textbf{Channel} & \multicolumn{2}{| l |}{\textbf{Experiment}} & \textbf{Mass range} & ${\cal L}$ \\
&&& \textbf{[TeV]} & \textbf{[fb$^{-1}$]} \\[1pt]
\hline
\hline
$pp \to S_{R,I} \to tt$ & ATLAS  & \cite{Aaboud:2018mjh} & [0.4;3] & 36.1 \\

\hline
$bb \to S_{R,I} \to tt$ & ATLAS & \cite{ATLAS-CONF-2016-104} & [0.4;1] & 13.2 \\
$pp \to S_{R,I} tt \to (tt)(tt)$ & ATLAS & \cite{Aaboud:2018xpj} & [0.4,1] & 36.1\\
\hline
$pp\to S^+ b \bar{t}\to t \bar{b} b \bar{t} $ &  ATLAS &\cite{Aad:2021xzu}& [0.2;2] & 139.1\\
\hline
\hline
$bb \to S_{R,I} \to bb$ & CMS8 & \cite{Khachatryan:2015tra} & [0.1;0.9] & 19.7\\
\hline
$gg \to S_{R,I}\to bb$ & CMS8 & \cite{Sirunyan:2018pas} & [0.325;1.2] & 19.7\\
\hline
$pp \to S_{R,I}\to bb$ & CMS & \cite{CMS-PAS-HIG-16-025} & [0.55;1.2] & 2.69\\
\hline
$bb \to S_{R,I}\to bb$ & CMS & \cite{Sirunyan:2018taj} & [0.3;1.3] & 35.7\\
\hline
$pp \to S_{R,I}\to gg$ & CMS & \cite{Sirunyan:2018xlo} & [0.5,8] & 27 \& 36\\
\hline
$pp \to S_{R,I}S_{R,I}\to (gg)(gg)$ & ATLAS & \cite{Aaboud:2017nmi} & [0.5,1.75] & 36.7\\
\hline
\end{tabular}
\caption{Data from direct searches included in the fit.
CMS8 means CMS at 8 TeV, the rest is at 13 TeV.}
\label{tab:directsearches1}
\end{small}
\end{center}
\end{table}

Since HEPfit cannot generate events by itself, the theoretical predictions for $\sigma\cdot\mathcal{B}$ must be provided with
other tools. We have used MadGraph \cite{Alwall:2014hca} to create tables with the values of $\sigma\cdot\mathcal{B}$ for different input choices of the parameters on which these observables depend: $\eta_U$, $\eta_D$, $(\nu_4+\nu_5)$ and $m_S$. These tables are read by HEPfit, which performs a linear interpolation to obtain the value of the observables at any point. The error of the linear interpolation is estimated to be  10\% or less for $\log(\sigma\cdot\mathcal{B})$. 

The experimental data are also provided in the form of tables, which compile the values of the 95\% upper limits on $\sigma\cdot\mathcal{B}$, as a function of the resonance mass. In these tables, HEPfit also performs a linear interpolation if needed.

In order to compare the theoretical results with the experimental data, we define the ratio of the theoretical prediction over the experimental upper limit. To this ratio we will assign a Gaussian distribution (restricted to positive values) centered at 0 such that the value 1 is excluded with a 95\% probability.

%%%%%%%%%%%%%%%%%%%%%%%%%%%%%%%%%%%%%%%%%%%%%%%%%%%%%
\subsection{Flavour constraints}
\label{sec:Flavourconstraints}
%%%%%%%%%%%%%%%%%%%%%%%%%%%%%%%%%%%%%%%%%%%%%%%%%%%%%

Regarding flavour transitions, we have included in the analysis the  $B^0_s-\bar{B}^0_s$ mixing and the decay $B^0_s\rightarrow \mu^+\mu^-$, which are the most constraining observables. The expressions for the relevant Wilson coefficients were taken from Ref.~\cite{Cheng:2015lsa}, where a complete one-loop calculation was performed. Only the charged scalars contribute to these processes at the one-loop level, and the corresponding amplitudes involve the parameters $\eta_U$ and $\eta_D$. The leading $\eta_U$ contributions are proportional to the top-quark mass, while the $\eta_D$ terms are weighted by the bottom-quark mass. Therefore, the transition amplitudes are mainly governed by the parameter $\eta_U$.

These observables strongly depend on the numerical values of the relevant quark mixings. We cannot use the standard determinations of the CKM entries, since they assume the validity of the SM. In order to obtain non-contaminated values, we perform a specific CKM fit using \texttt{HEPfit}, which only includes observables that are not affected by the coloured scalars. 
The inputs of this CKM fit, shown in Table \ref{tab:CKMinputs}, are taken from the PDG 2020 \cite{Zyla:2020zbs} except $|V_{cs}|$, for which we have adopted the more recent result of Ref.~\cite{Chakraborty:2021qav}. The value of $|V_{ud}|$ 
has been considerably shifted with respect to the one quoted in the PDG 2018 \cite{Tanabashi:2018oca}. Combined with the reduction of its uncertainty, this generates a violation of unitarity in the first row of the CKM matrix at the $3\sigma$ level. 
Since there is no full consensus in the community, we
have adopted a conservative attitude, increasing
the error of $|V_{ud}|$ by a factor $2.4$ so that CKM unitarity is recovered at 2$\sigma$, 
without modifying any central values. The results of this CKM fit, shown in Table \ref{tab:WolfResult}, are then used as inputs in our analysis of the MW parameter space.

\begin{table}[h!]
\centering
\begin{tabular}{c|c|p{70 mm}}
\textbf{CKM Parameter} & \textbf{Input Value} & \multicolumn{1}{|c}{\textbf{Source}}  \\
\hline
 $|V_{ud}|$ & $0.97370\pm 0.00033$ & Superallowed $0^+\to0^+$ nuclear $\beta$ \newline decays \\
\hline
 $|V_{us}|$ & $0.2245 \pm 0.0008$  & Kaon and pion semileptonic and leptonic decays \\
\hline
 $|V_{cd}|$ & $0.221 \pm  0.004$  & \multirow{2}{70 mm}{Semileptonic and leptonic charm decays} \\
\cline{1-2}
 $|V_{cs}|$ & $0.966 \pm 0.008$  &  \\
\hline
 $|V_{cb}|$ & $(41.0 \pm 1.4)\times 10^{-3} $   & \multirow{2}{70 mm}{Exclusive and inclusive semileptonic decays of $B$ mesons}    \\
\cline{1-2}
 $|V_{ub}|$ & $(3.82 \pm 0.24)\times 10^{-3} $   &   \\
\hline
 $|V_{td}/V_{ts}|$ & $0.205  \pm  0.006$   & $B^0-\overline{B}^0$ mixing  \\
\hline
\end{tabular}

\caption{Inputs \cite{Zyla:2020zbs,Chakraborty:2021qav} used for the CKM fit and processes from which they are obtained. The error of $|V_{ud}|$ has been increased by a factor 2.4.}
\label{tab:CKMinputs}
\end{table}

\begin{table}[h!]
\centering
\begin{tabular}{c|c|c|c|c|c}
    \multirow{2}{*}{ } & \multirow{2}{*}{Value} & \multicolumn{4}{c}{Correlation}  \\\cline{3-6}
    & & $ \lambda $ & $A$ & $\overline{\rho}$ & $ \overline{\eta}$\\ \hline
    $\lambda$ & $0.22518 \pm 0.00069$        & 1 & $-0.18$ & 0.08 & $-0.084$\\
    $A$ & $0.808  \pm 0.028$               & $-0.18$  & 1  & $-0.22$  & $-0.41$\\
    $ \overline{\rho}$ & $0.181\pm 0.026$  & 0.08 & $-0.22$ & 1 & 0.039  \\
    $ \overline{\eta}$  & $0.360\pm 0.029$ & $-0.084$ & $-0.41$ & 0.039 & 1 \\
\end{tabular}

\caption{Wolfenstein parameters obtained from a fit to
the CKM entries in Table~\ref{tab:CKMinputs}.}
\label{tab:WolfResult}
\end{table}

%%%%%%%%%%%%%%%%%%%%%%%%%%%%%%%%%%%%%%%%%%%%%%%%%%%%%
\subsection{Electroweak precision data}
\label{sec:EWpd}
%%%%%%%%%%%%%%%%%%%%%%%%%%%%%%%%%%%%%%%%%%%%%%%%%%%%%

In this work, we are studying the most general CP-conserving MW model, so we have not imposed any additional assumptions such as  Custodial Symmetry. Therefore, the coloured scalars generate non-vanishing contributions to the oblique parameters \cite{Peskin:1990zt,Peskin:1991sw} that can be used to constrain the parameter space of the model. 
The standard bounds on $S$, $T$ and $U$ \cite{Zyla:2020zbs} include in the global fit 
the ratio $R_b= \Gamma (Z\to b \bar{b})/\Gamma (Z\to \mathrm{hadrons})$, which incorporates quantum corrections to $\Gamma(Z\to b \bar b)$ that are enhanced by the large value of the top quark mass \cite{Bernabeu:1987me,Bernabeu:1990ws}. Since the ratio $R_b$ receives also sizeable corrections from the additional scalars, we cannot use these bounds. Similarly to what has been done before for the CKM matrix, we first determine the oblique parameters, performing a combined fit with \texttt{HEPfit} of electroweak precision observables without $R_b$ \cite{Eberhardt:2020dat}, and then use the resulting values as inputs in our analysis of the MW model. The values of the oblique parameters that we have obtained from this fit are shown in Table~\ref{tab:STU}.

\begin{table}
\begin{center}
\begin{tabular}{c|c|c|c|c}
    \multirow{2}{*}{ } & \multirow{2}{*}{Value} & \multicolumn{3}{c}{Correlation}  \\\cline{3-5}
    & & $ S $ & $ T $ & $ U $ \\ \hline
    $ S $ & 0.093$\pm$ 0.101 & 1.00  &  0.86  &  -0.54\\
     $ T $ & 0.111$\pm$0.116 & 0.86  &  1.00  &  -0.83\\
     $ U $ & -0.016$\pm$0.088 & -0.54 &  -0.83 &  1.00\\
\end{tabular}
\end{center}
\caption{Results for the fit of the oblique parameters $S$, $T$ and $U$, excluding the information from $R_b$.}
\label{tab:STU}
\end{table}

The theoretical predictions for $S$, $T$ and $U$ in the MW model have been taken from Ref.~\cite{Burgess:2009wm}. The oblique parameters depend on the masses of the new scalars and, therefore, provide important constraints on $m_S$, $\nu_1$, $\nu_2$ and $\nu_3$. Once, the oblique parameters have been fixed, we also include $R_b$ in the fit to the MW parameter space. The theoretical prediction of this observable, including the QCD corrections, has been taken from Ref.~\cite{Degrassi:2010ne}. The ratio $R_b$ will basically constrain $\eta_U$ because, like for the flavour observables, the new physics contribution comes from 
one-loop diagrams involving virtual charged scalars, coupled to heavy quarks.

%%%%%%%%%%%%%%%%%%%%%%%%%%%%%%%%%%%%%%%%%%%%%%%%%%%%%
\section{Results}
\label{sec:results}
%%%%%%%%%%%%%%%%%%%%%%%%%%%%%%%%%%%%%%%%%%%%%%%%%%%%%
\subsection{Theoretical constraints}

\begin{figure}[h!]
    \centering
    \includegraphics[scale=0.5]{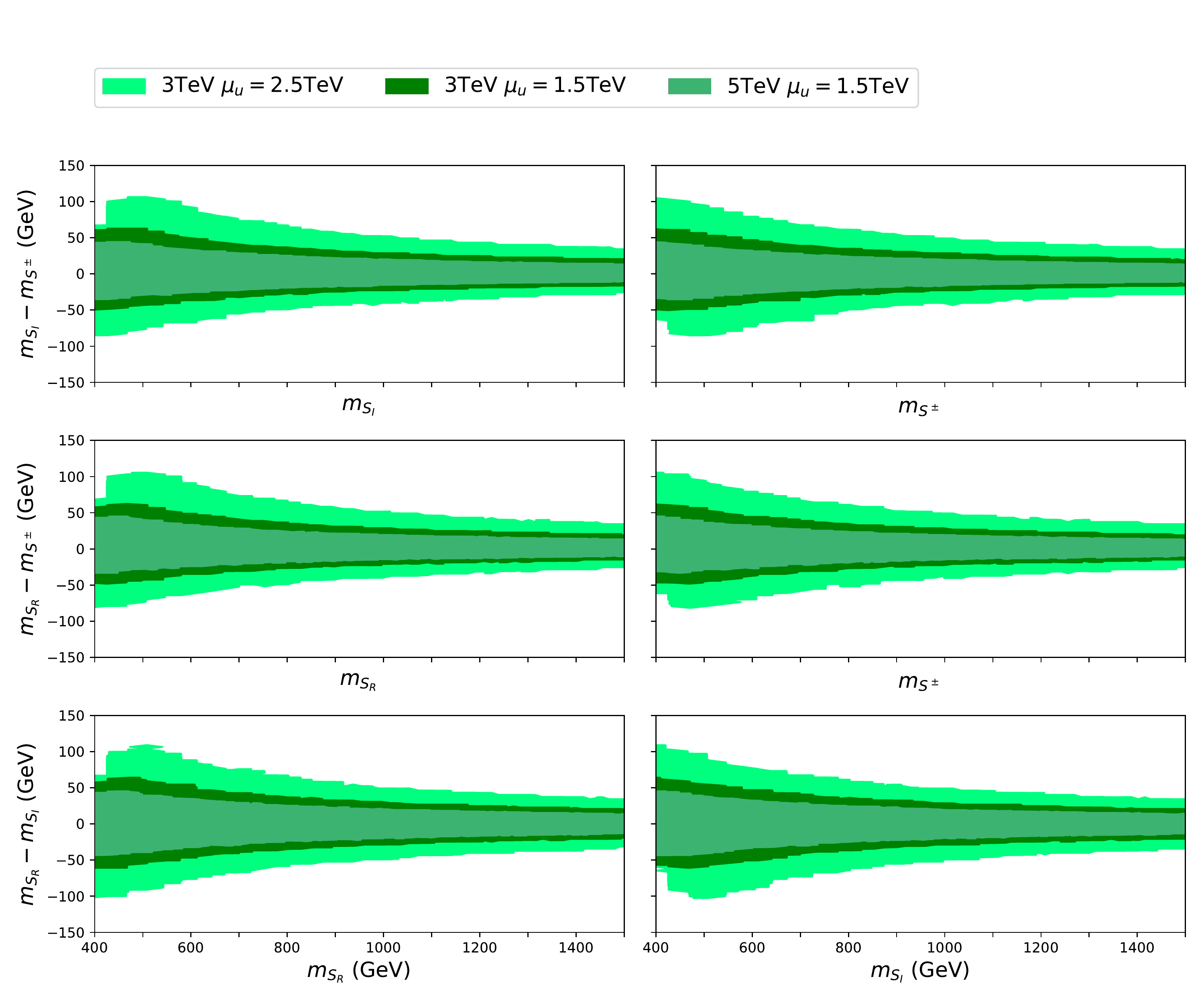}
    \caption{Theoretical constraints on the mass splittings, as functions of the scalar masses. The different coloured regions correspond to imposing RG stability up to an UV mass scale of 3 or 5 TeV, and imposing perturbative unitarity conditions above 2.5 TeV or 1.5 TeV.}
    \label{fig:Theo_delta_mass}
\end{figure}

As mentioned before, we have imposed that this model should
satisfy perturbative unitarity and be well defined up to, at least, 3 TeV. These conditions generate bounds on all the parameters of the scalar potential that, in some cases, can be 
stronger than the ones obtained from experimental data. For instance, the quartic couplings are not constrained experimentally but the theoretical requirements imply that none
of them can be, in modulus, higher than 6.  
The theoretical constraints can be translated to physical observables like the mass differences, which depend on these parameters. In Fig.~\ref{fig:Theo_delta_mass} we show the theoretical constraints on the scalar mass differences for
two choices of the UV scale, 3 TeV and 5 TeV. Obviously, requiring the model to be valid up to a higher scale results in stronger bounds. The figure exhibits also the importance of perturbative unitarity, showing how the limits 
get weakened if the scale above which we impose it is chosen to be higher.

The large impact of these theoretical constraints can be better appreciated in Fig.~\ref{fig:All_theo}, which displays the bounds they impose on the $\mu_n$ and $\nu_n$ potential parameters, in the form of two-dimensional correlated plots, compared with the final results from the global fit, including all constraints, that we will later discuss.

\subsection{Experimental constraints}
\begin{figure}[h!]
    \centering
    \includegraphics[scale=1.]{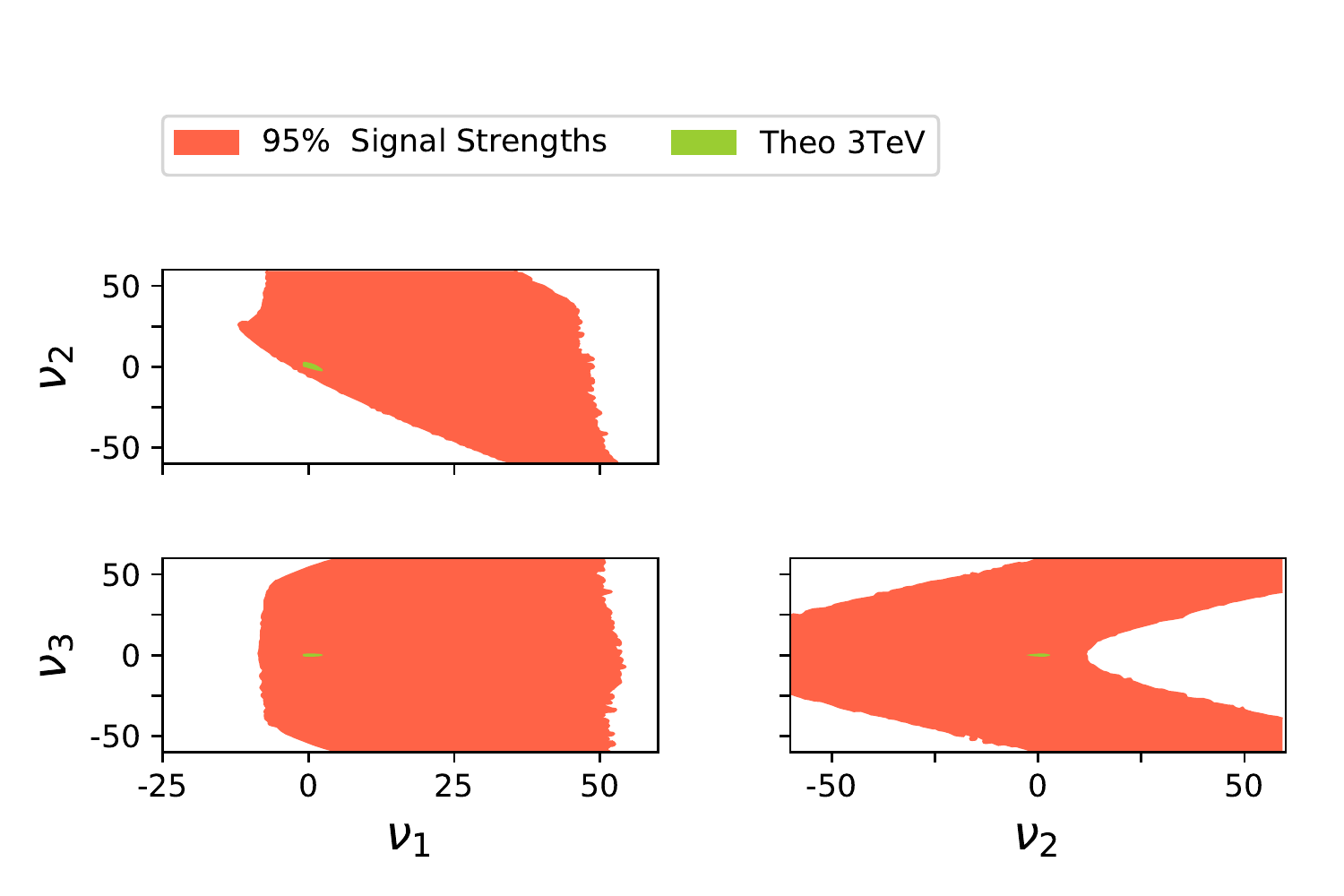}
    \caption{Two-dimensional constraints on $\nu_1$, $\nu_2$ and $\nu_3$ from Higgs signal strengths (red), compared with the theoretically allowed regions (green), assuming RG stability up to 3 TeV.}
    \label{fig:SignalStrengths}
\end{figure}

As we have shown before, using theoretical arguments we can constrain the parameter space of all the new quartic couplings that appear in the scalar potential. In this section we will use experimental measurements in order to constrain the masses of the coloured scalars and their Yukawa couplings. But first of all let us consider the observables that constrain also the parameters of the potential and let us compare the results with the ones obtained from theoretical requirements. 

In Fig.~\ref{fig:SignalStrengths} we show the constraints 
on $\nu_1$, $\nu_2$ and $\nu_3$ emerging from the Higgs signal strengths and compare them with the theoretically allowed regions, imposing RG stability up to 3 TeV.
Clearly the theoretical restrictions are much stronger than the Higgs constraints so, a priori, one could think that the latter will be irrelevant in the global fit. However, the addition of these experimental observables will have some effect on the global fit, as we will show later. The fact that the constraints from the Higgs signal strengths, alone, are so weak is an expected behaviour, given that the coloured scalars only start to contribute to them at the one-loop level.

\begin{figure}[t]
    \centering
    \includegraphics[scale=0.8]{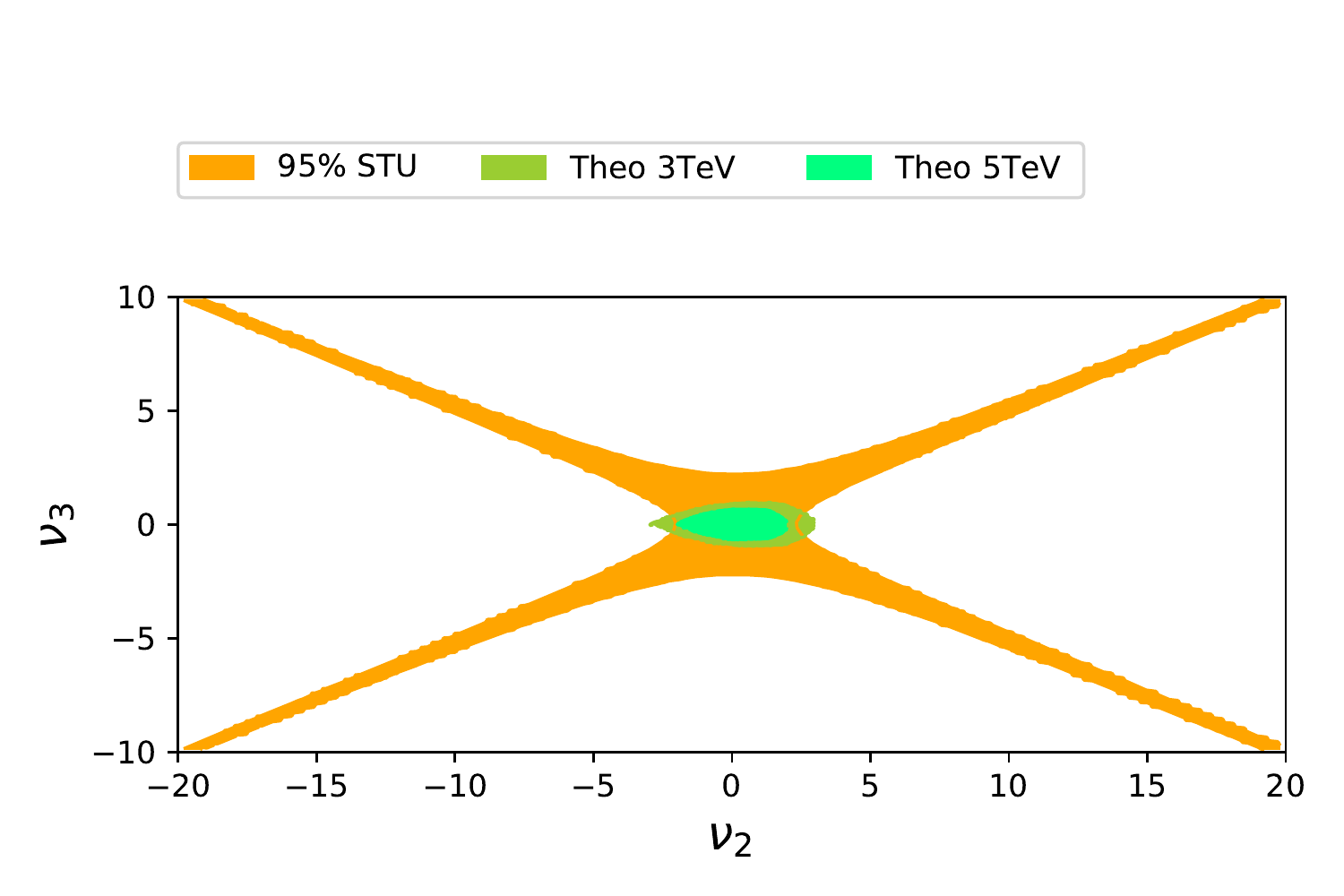}
    \caption{Allowed two-dimensional regions of $\nu_2$ and $\nu_3$ from the oblique parameters (orange, 95\% probability) and from theoretical constraints (green) with the NLO+ approximation. The darker and lighter green areas correspond to requiring RG stability up to 3 and 5 TeV, respectively.} 
    \label{fig:STU}
\end{figure}

The oblique parameters constrain the scalar mass splittings, bounding the combinations of quartic couplings $\nu_2\pm 2\nu_3$ that appear in Eq.~\eqref{eq:msplit}. This is clearly seen in the narrow allowed regions (orange) displayed in Fig.~\ref{fig:STU}. At large values of $\nu_{2,3}$, the oblique parameters impose that $\nu_2\approx\pm 2\nu_3$ in order to reduce the splitting between the charged scalar and either the CP-odd or CP-even neutral scalars. The oblique bounds are relevant even when we consider the theoretical constraints up to 5 TeV, but they are specially important when we only require RG stability up to 3 TeV. Indeed, combining both constraints we can obtain a harder restriction on these parameters for the 3 TeV case.

\begin{figure}
    \centering
    \includegraphics[scale=1.0]{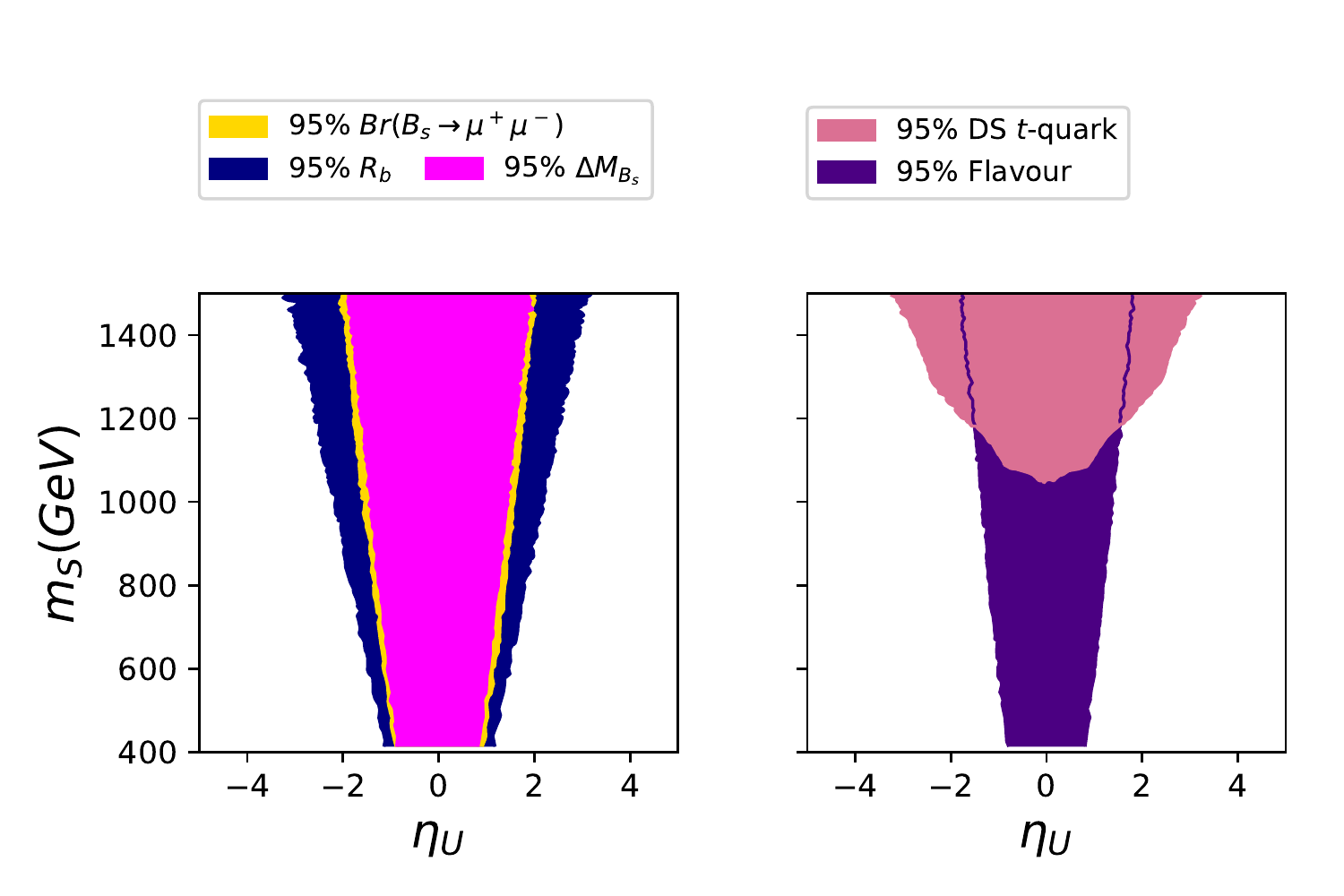}
    \caption{Experimental constraints on the $m_S$-$\eta_U$ plane. Left panel: allowed regions at 95\% probability obtained from $R_b$ (blue), $\text{Br}(B_s\rightarrow \mu^+\mu^-)$ (yellow) and $\Delta M_{B_s}$ (magenta). Right panel: combined flavour constraints, compared with the limits from direct searches including top quarks, at 95\% probability.}
    \label{fig:Flavour_Rb_DS}
    \end{figure}

The absolute mass scale $m_S$ is constrained by the electroweak ratio $R_b$, which is also sensitive to the Yukawa alignment parameter $\eta_U$. The $m_S$-$\eta_U$ plane is also constrained by the two flavour observables that we have considered: $B^0_s-\bar{B}^0_s$ mixing and $\text{Br}(B_s\rightarrow \mu^+\mu^-)$.
The left panel of Fig.~\ref{fig:Flavour_Rb_DS} displays the regions allowed by each of these observables, individually, at 95\% probability. The mass difference between $B^0_s$ and $\bar{B}^0_s$ turns out to provide the strongest bounds, although they are quite similar to the ones from $\text{Br}(B_s\rightarrow \mu^+\mu^-)$. Furthermore, since the three observables exhibit a similar dependence on these parameters, the combined constraints obtained with the three observables together are not much better than the ones given by just $\Delta M_{B_s}$ or $\text{Br}(B_s\rightarrow \mu^+\mu^-)$.

The $m_S$-$\eta_U$ plane is also constrained by the direct searches, as shown in the right panel of Fig.~\ref{fig:Flavour_Rb_DS}. In this figure we have only included searches with at least one top quark in the final state because those are the channels generating the most interesting constraints. Since the figure does not include any observable constraining the scalar mass differences, we have set these mass splittings to zero.
Doing so, we are able to totally exclude values of $m_S$ (equal in this case to the physical scalar masses) smaller than 1.1~TeV, for any value of the other parameters, with a probability of 95\%. The plot shows also the additional constraints obtained on this plane from flavour observables.

\begin{figure}
    \centering
    \includegraphics[scale=0.8]{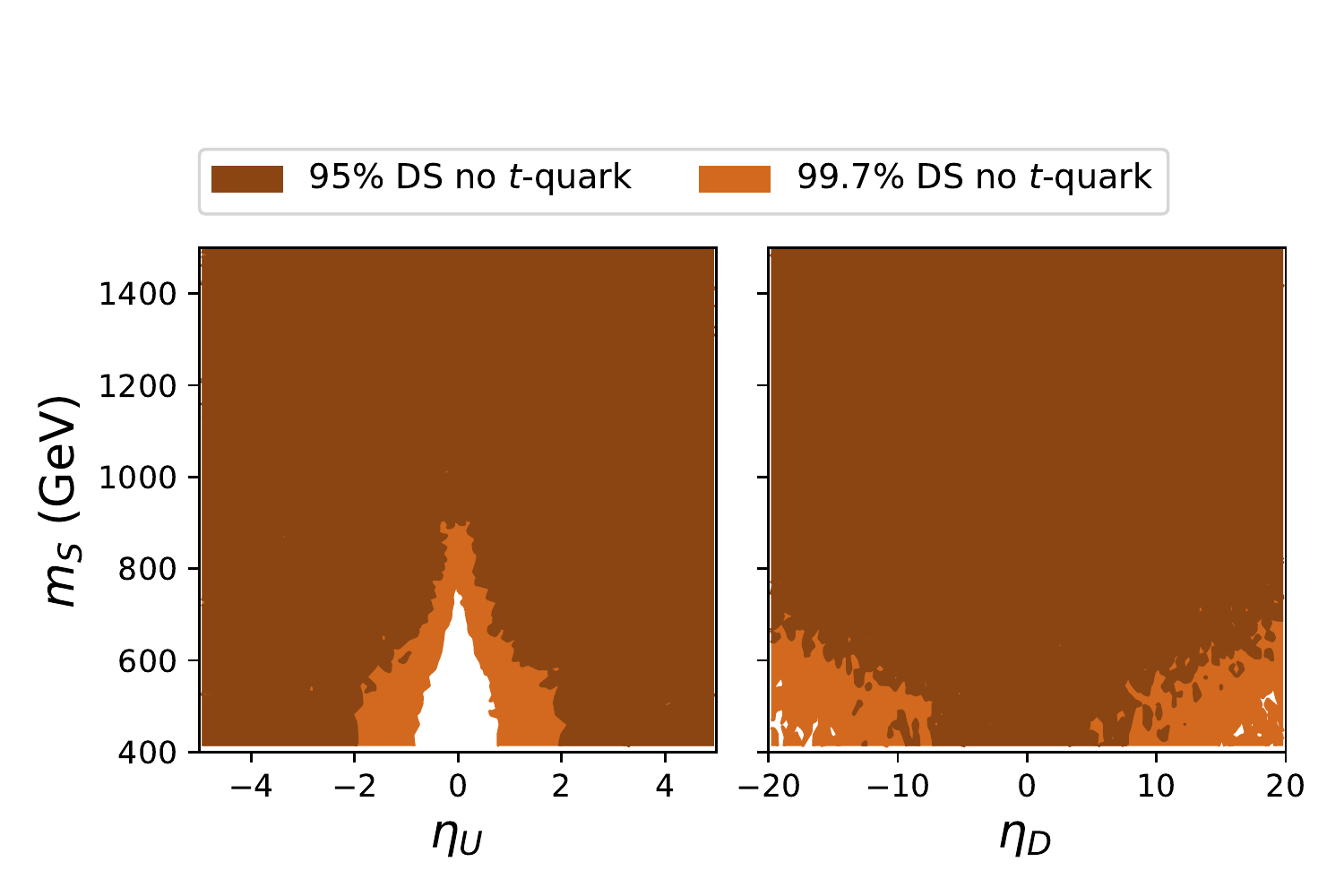}
    \caption{Constraints on the $m_S$-$\eta_U$ (left) and $m_S$-$\eta_D$ (right) planes,
    obtained from direct searches in channels without top quarks in the final state.}
    \label{fig:Ds_No_Top}
    \end{figure}

In Fig.~\ref{fig:Ds_No_Top} we display the analogous constraints emerging from direct searches in channels without top quarks in the final state. Relevant bounds are only obtained at small values of $\eta_U$ or for high values of $\eta_D$. This is to be expected because those are the parameter regions where the scalar decays into top quarks are suppressed and, therefore, the branching ratios for the other channels are increased. The decay channels to top quarks are of course the dominant ones outside these particular regions of the parameter space, and decay modes without top quarks become irrelevant for large values of $\eta_U$ or small values of $\eta_D$.

Similar features are found in the $m_S$-$\eta_D$ plane, where the bound $m_S > 1.1$~TeV is also obtained with a 95\% probability from direct searches in channels with top-quark production. This could be, in principle, a bit surprising because one could naively expect that for very small values of $\eta_U$ those searches should not generate any constraint. 
This is true for the channels in which a neutral scalar decays to a $ t \bar{t}$ pair, but not for a charged scalar decaying to $ t \bar{b}$, a process which also depends on $\eta_D$. Indeed, in the right panel of Fig.~\ref{fig:Flavour_Rb_DS}, one observes that the lower bound on $m_S$ decreases when $\eta_U$ approaches zero. This is because all channels with neutral scalars become irrelevant in that limit, but the information from charged-scalar channels is still good enough to generate a quite strong constraint.
Therefore, as long as $\eta_U$ and $\eta_D$ are not both extremely close to zero, we obtain good constraints on $m_S$. Since the MW model is motivated by MFV, the particular region where the two quark Yukawa couplings are both zero does not seem to have much theoretical interest.

\subsection{All constraints}

\begin{figure}[h!]
    \centering
    \includegraphics[scale=0.5]{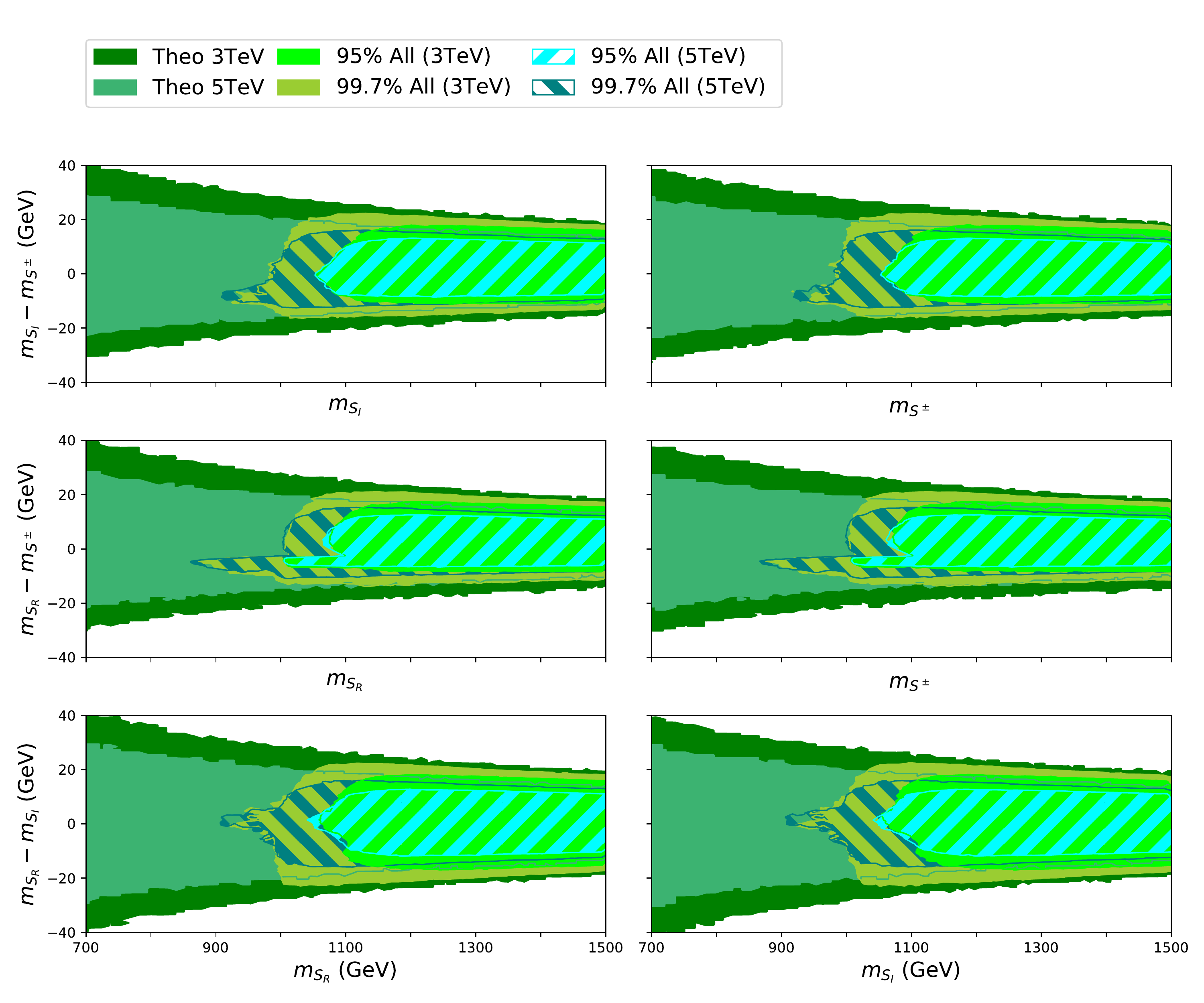}
    \caption{Comparison of theoretical and global  constraints on the mass splittings, as a function of the different scalar masses, imposing RG stability and perturbative unitarity up to 3 and 5 TeV. For the global fit we plot the 95\% and 99.7\% regions.}
    \label{fig:delta_mass}
\end{figure}

Once we have analysed the constraints emerging from individual observables, we can combine all of them into a global fit. The first interesting result is that we are able to find lower limits for the physical masses of our scalars and constraints on their mass splittings, as shown in Fig.~\ref{fig:delta_mass}. Note that the mass differences between the different scalars are now restricted to be smaller than 25 GeV with a 99.7\% probability, if we impose RG stability and perturbative unitarity up to 3 TeV, and to be smaller than 20 GeV, with the same probability, when we go up to 5 TeV. This justifies a posteriori our approximation of not including in the analysis of direct searches the decays of the coloured scalars into another scalar plus a weak boson.

\begin{figure}[h!]
    \centering
    \includegraphics[scale=1.0]{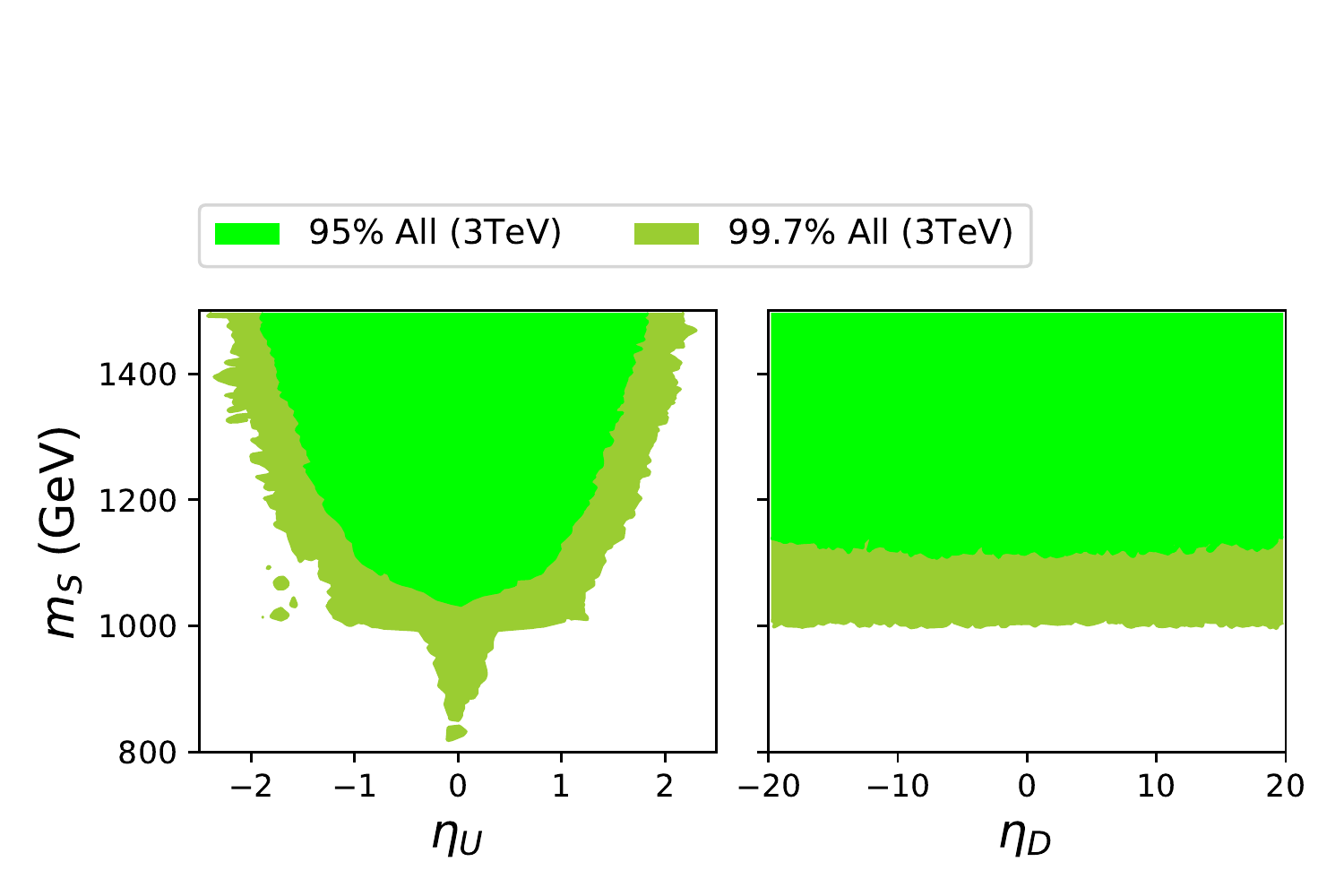}
    \caption{Allowed regions on the $m_S$-$\eta_U$ (left) and $m_S$-$\eta_D$ (right) planes from the global fit, at 95\% and 99.7\% probability, imposing RG stability up to 3 TeV and perturbative unitarity at NLO+. Quite similar results are obtained, requiring  RG stability up to only 1 TeV.}
    \label{fig:EtaU_EtaD_vs_mS_All}
\end{figure}

Another interesting result is shown in Fig.~\ref{fig:EtaU_EtaD_vs_mS_All}, where we plot again the
constraints on the $m_S$-$\eta_U$ and $m_S$-$\eta_D$ planes. We can clearly see how $\eta_D$ is not constrained. On the other hand, $|\eta_U|$ cannot take values higher than 1.8, for masses of the scalars smaller than 1.5 TeV, within a probability of 95\%. 
The constraint on the scalar mass scale is mainly coming from direct searches including decays to top quarks. This is why for values of $\eta_U$ close to zero the lower bounds on $m_S$ are weaker. However, as mentioned before, even for very small values of $|\eta_U|$ the constraints remain strong. We have also performed some global fits in which we set $\eta_U=0$ or $\eta_D=0$. For the first case we found that the limits on the scalar masses are roughly 100 GeV smaller, while for the second we found the same result than for the global fit in which we vary both parameters. Therefore, even in the hypothetical case that some symmetry would force one of the Yukawa couplings to vanish, one would still find important constraints on the scalar masses.

\begin{figure}[h!]
    \centering
    \includegraphics[scale=0.25]{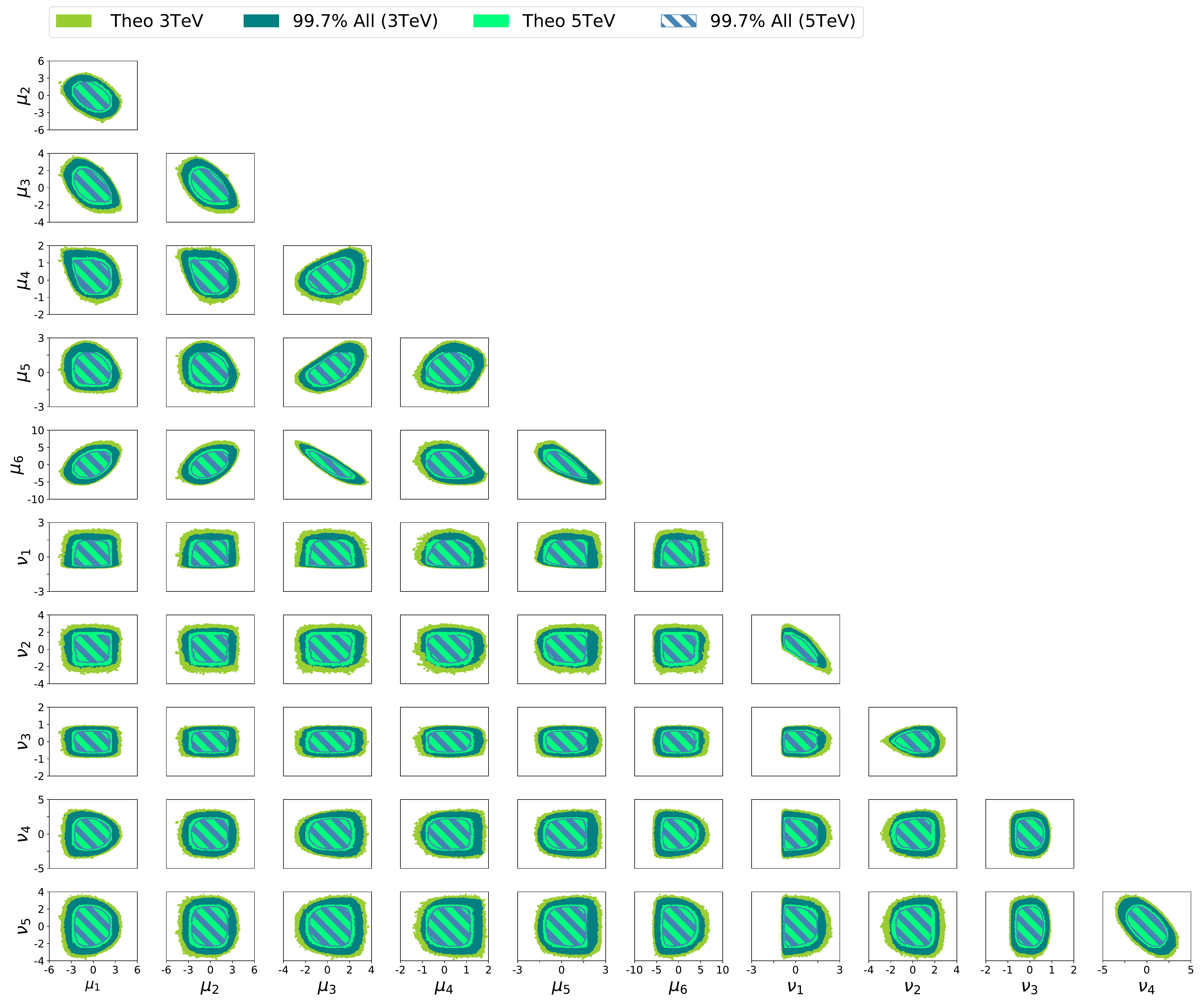}
    \caption{Two-dimensional correlations among the quartic couplings of the scalar potential. The
    allowed (99.7\% probability) regions from the global fit are compared with the results obtained with only theoretical constraints, imposing perturbative unitarity and RG stability up to 3 and 5 TeV.}
    \label{fig:All_theo}
\end{figure}

The two-dimensional correlations among the quartic potential parameters are displayed in Fig.~\ref{fig:All_theo}, which compares the limits emerging from the global fit with those obtained with only theoretical constraints, imposing RG stability and perturbative unitarity up to 3 and 5 TeV.
The allowed regions from theoretical constraints must be understood as having 100\% probability because they correspond to a discrete condition: a given point of the parameter space is either allowed or not.
This introduces a small difference when comparing the theoretical and global limits, since the latter ones refer to a slightly smaller probability. Taking this into account, we can still see that $\nu_1$ and $\nu_2$ are slightly more constrained in the global fit, if RG stability is only imposed up to 3 TeV. Nevertheless, for the other observables the allowed regions remain almost invariant when including the experimental information. This is not a surprise because the measured observables depend on $\nu_1$, $\nu_2$ and $\nu_3$.
The improvements introduced by the global fit are a consequence of adding the oblique parameters and, specially, the Higgs signal strengths.
In order to check this, we performed a fit with only theory and the Higgs signal strengths and the results for $\nu_1$ and $\nu_2 $ were basically identical to the ones of the global fit.
Indeed, although the Higgs signal strengths alone do not produce strong limits,  combining them with the theoretical constraints, which reduce the allowed range of $\nu_3$, results in very good bounds on $\nu_1$ and $\nu_2$. 

If RG stability is imposed up to 5 TeV, the current experimental measurements have a quite small impact in the allowed regions in Fig.~\ref{fig:All_theo}, which are mainly governed by the much stronger theoretical constraints.

Finally, in Table~\ref{tab:munuglobal} we present the marginalised allowed ranges for the parameters of the potential, from the global fit, with a probability of 95\%. As can be seen there, none of the quartic couplings can be, in modulus, higher than 5 within this probability.

\begin{table}
\begin{center}
\begin{small}
\begin{tabular}{|c|c|c|}
\hline
\textbf{Variables} & \textbf{All 3TeV}  & \textbf{All 5TeV}  \\
\hline
$\mu_1$ & (-2.9, 2.3) & (-1.3, 1.5) \\
\hline
$\mu_2$ & (-2.7, 2.6) & (-1.8, 1.8) \\
\hline
$\mu_3$ & (-1.6, 2.4) & (-1.0, 1.6) \\
\hline
$\mu_4$ & (-0.60, 1.35) & (-0.44, 0.90) \\
\hline
$\mu_5$ & (-1.0, 1.9) & (-0.7, 1.3) \\
\hline
$\mu_6$ & (-4.6, 3.5) & (-2.9, 2.4) \\
\hline
$\nu_1$ & (-0.8, 1.4) & (-0.66, 0.97) \\
\hline
$\nu_2$ & (-1.2, 2.0) & (-0.95, 1.47) \\
\hline
$\nu_3$ & (-0.65, 0.70) & (-0.45, 0.49) \\
\hline
$\nu_4$ & (-2.4, 2.4) & (-1.7, 1.6) \\
\hline
$\nu_5$ & (-1.6, 1.5) & (-1.7, 1.6) \\
\hline
\end{tabular}
\caption{Allowed ranges for the quartic potential parameters from the global fit, at 95\% probability.}
\label{tab:munuglobal}
\end{small}
\end{center}
\end{table}

%%%%%%%%%%%%%%%%%%%%%%%%%%%%%%%%%%%%%%%%%%%%%%%%%%%%%
\section{Summary}
\label{sec:summary}
%%%%%%%%%%%%%%%%%%%%%%%%%%%%%%%%%%%%%%%%%%%%%%%%%%%%%

In this work, we have presented the first global fit of the MW model. We have combined the more relevant theoretical and experimental constraints, including flavour and electroweak precision observables, Higgs signal strengths and direct searches.
The theoretical constraints have a quite large impact on the scalar parameter space, specially when RG stability is imposed up to 5 TeV, providing bounds on all the scalar potential parameters. The current experimental information on the Higgs signal strengths and the oblique parameters gives an improved sensitivity to $\nu_1$, $\nu_2$ and $\nu_3$, which becomes relatively more important if RG stability is only imposed up to 3 TeV. This allows for more stringent constraints on the scalar mass splittings.

The flavour observables and $R_b$ constrain the $m_S$-$\eta_U$ plane, but all of them in the same direction. The strongest limits come from $\Delta M_{B_s}$ and $\text{Br}(B_s\rightarrow \mu^+\mu^-)$, which combined, require that 
\begin{equation}
|\eta_U|\, <\, 1.8
\end{equation}
for scalar masses smaller than 1.5 TeV, within a probability of 95\%. 

A quite strong bound on the absolute mass scale $m_S$ emerges from the LHC data on direct searches for new scalars. The more sensitive channels, which involve the production of top quarks in the final state, imply that the masses of all coloured scalars must satisfy the bound
\begin{equation}
    m_{S^\pm} ,m_{S_R}, m_{S_I}\, >\, 1.05\;\mathrm{TeV}
\end{equation}
for any value of the other parameters, with a 95\% probability.
The global fit also restricts the scalar mass splittings to be smaller than 20 GeV, with a 95\% probability, even when RG stability is only imposed up to 3 TeV.

As shown in Ref.~\cite{Miralles:2019uzg}, even for tiny values of $|\eta_U| > 10^{-7}$, one still finds a strong lower bound on the scalar masses, provided $|\eta_D| > 10^{-5}$. These bounds can only be avoided for fermiophobic scalars with $\eta_U\approx\eta_D\approx 0$, such that their decay into a fermion-antifermion pair is highly suppressed. In that case, they would have a completely different experimental signature, since they would either decay into a lighter coloured scalar and a gauge boson or would behave as strongly-interacting long-lived particles. Although fermiophobic scalars are not compelling from the MFV point of view of the MW model, their phenomenology could be interesting by itself, but requires a more specific analysis that we postpone to future works. 

The forthcoming Run3 of the LHC and its subsequent high-luminosity phase will provide much larger data samples, substantially increasing the sensitivity to coloured scalar particles. Mass scales up to 1.3 TeV seem to be reachable with the expected luminosity of 3 $\text{ab}^{-1}$. The constraints on those potential parameters not related to scalar masses will remain, however, largely dependent on theoretical considerations, unless a real discovery of a colourful scalar state emerges from the data.

\section*{Acknowledgements}

This work has been supported in part by the Spanish Government and ERDF funds from the EU Commission [grant FPA2017-84445-P], by the Generalitat Valenciana [grant Prometeo/2017/053] and by the COST Action CA16201 PARTICLEFACE.
The work of V.M. is supported by the FPU doctoral contract FPU16/0191, funded by the Spanish Ministry of Universities.

\bibliographystyle{JHEP}
\bibliography{Coloured_Scalars_Fits}

\end{document}